\documentclass[pra,preprint,showpacs]{revtex4}
\usepackage{graphicx}
\begin{document}
\title{Independent-subsystem interpretation of the double photoionization 
of pyrene and coronene
} 
\author{D. L. Huber and R. Wehlitz$^{\ast}$} 
\affiliation{
University of Wisconsin--Madison, Madison, Wisconsin 53706, USA\\
$^\ast$e-mail: rwehlitz@gmail.com
}
\date{\today}

%
%
\begin{abstract}
It is shown that the M$^{2+}$ ion yield in the double photoionization 
of the aromatic hydrocarbons, pyrene and coronene, can be expressed as 
a superposition of a contribution from a resonance involving carbon 
atoms on the perimeter and coherent contributions from carbon atoms 
inside the perimeter. In the case of pyrene, the two interior atoms 
are associated with a resonance peak at 10 eV and linear behavior above 
75 eV.  The resonance peak is an optically excited state of the interior
carbon pair.  
The linear behavior arises from the coherent emission of two electrons 
with equal energy and opposite momenta, as occurs in pyrrole.  Coronene has 
a low energy peak along with a pairing resonance, however, the linear
region as in the case of pyrene is absent.  The resonance is associated with 
the atoms on the perimeter. It is proposed that the quasi-independence of the 
contributions from the perimeter and interior atoms is related to H\"uckel's
Rule for the stability of aromatic hydrocarbons.
\end{abstract}
\maketitle

\section{Introduction}
The photon-energy dependence of the simultaneous removal of two electrons 
from an atom or molecule by a single photon has been studied for several cases
(see, e.g., \cite{Wehl10}). In recent publications \cite{Wehl14,Hart13} 
the double photoionization (DPI) results for the aromatic hydrocarbons pyrene 
and coronene using monochromatized synchrotron radiation over a wide range of 
energies have been reported. The general goal of these investigations is to 
elucidate the role of the physical molecular structure on the DPI process. 
A unique feature among the hydrocarbons studied using DPI is that pyrene 
and coronene have carbon atoms located inside their molecular 
perimeter as well as on their perimeter \cite{Wehl16}. The chemical formula of 
pyrene is C$_{16}$H$_{10}$ with 14 carbon atoms on the perimeter and 
two carbon atoms in the interior, positioned as nearest-neighbors. Coronene has 
the formula C$_{24}$H$_{12}$ with 18 carbon atoms on the perimeter and six 
carbon atoms in the interior that are arranged in a benzene-like ring.  The 
focus in this paper is the role played by carbon atoms inside 
the perimeters of the two molecules on DPI. We will discuss that the 
perimeter carbon atoms and the interior carbon atoms can be regarded as 
independent subsystems in the DPI process. The current analysis of the 
experimental data was undertaken in light of the previous theoretical 
studies of DPI in aromatic hydrocarbons \cite{Hube14,Hube18}.
  
Our results are presented in plots of the DPI ratio vs excess energy 
relative to the DPI threshold. Here, the DPI ratio is defined by the expression 
\begin{equation}
R = M^{2+} / M_{\rm tot} - K,
\label{ratio}
\end{equation}
where $M^{2+}$ and $M_{\rm tot}$ denote the yields 
of doubly charged and singly plus doubly charged parent ions, respectively. 
$K$ is the knock-out contribution \cite{Wehl16} to the DPI process, in 
which the first electron on its way out of the molecules knocks out a 
second electron. The contribution from the knock-out mechanism is modelled
by the double-to-total photoionization ratio of helium, which is justified
for photon energies as low as in this study. We begin our analysis with 
pyrene and coronene, and will then test our model of independent subsystems
on anthracene.  

Note that this preprint was updated because of a mistake in the data analysis.
The additional hump in the 80 to 130 eV range shown in Fig.\ 3 of the previous 
version does not exist. Nevertheless, the basic conclusions presented in 
this preprint are still the same.



\section{Pyrene}

\begin{figure}[htb]
\includegraphics[width=8.8cm,clip]{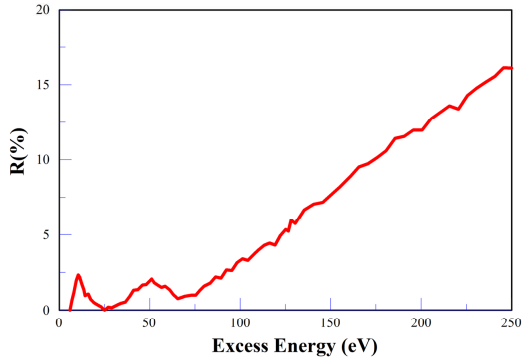}
\caption{
Relative DPI yield $R$ as defined in Eq. (\ref{ratio})
in pyrene as a function of excess 
energy. The solid line connects the data points. Corresponding error bars 
are given in \cite{Wehl16}. See text for details.}
\label{one}
\end{figure}

In pyrene, after subtracting the contribution from the knock-out mechanism 
to the DPI ratio as discussed in \cite{Hart13}, there are three regions in 
the photon energy spectrum with distinctly different behaviors in the DPI:  
(i) a low-energy region, with a comparatively sharp resonance peak at 10 eV 
above the DPI threshold, (ii) a mid-range region with a 25 eV onset and a 
broad peak at 51 eV, and (iii) a quasi-linear region extending to more than 
250 eV above threshold (Fig.\ \ref{one}).  We associate the low-energy 
and linear regions with the two carbon atoms inside the perimeter of 
the molecule, while the mid-range region is linked to the atoms on the 
perimeter.
  
In the 10-eV resonance, the energy of the photon that has been absorbed 
is equal to the sum of the resonance energy and the energy required to promote 
two electrons to the DPI threshold. A theory for the origin of the low-energy 
resonance in pyrene is discussed in \cite{Hube18}. The approach is based on 
the Hubbard model \cite{Hubb63}, sometimes referred to as the 
Pariser-Parr-Pople model. The optical absorption associated with the 
two-site Hubbard Hamiltonian characterizes a process where a $\pi$ electron 
is transferred to the site that is already occupied by a $\pi$ electron 
with opposite spin.  It consists of a single peak at the energy 
$E_{OA}$ where \cite{Maes06}
\begin{equation}
E_{OA} = U/2 +[ (U/2)^2 +4t^2]^{1/2}.
\end{equation}
Here $t$ denotes the transfer integral characterizing the interaction 
between $\pi$ electrons on nearest-neighbor sites and $U$ is the 
electrostatic interaction between two $\pi$ electrons on the same carbon 
site.  In order to calculate the location of the DPI peak, we take parameter 
values appropriate for benzene: $U = 10$ eV and $t = 2.5$ eV \cite{Burs98} 
and obtain the result $E_{OA} = 12$ eV, in reasonable agreement with 
experiment.

An alternative interpretation of the low-energy resonances in pyrene and 
coronene is outlined in \cite{Wehl14}. As discussed there, the peak is 
identified as a pairing resonance having a de Broglie wavelength of 
2.8 \AA, i.e.\ twice the C-C distance.  

The linear behavior of the DPI shown in Fig.\ \ref{one} resembles the 
linear behavior in pyrrole and the related molecules furan and 
selenophene \cite{Wehl16}. These molecules consist of a five-member ring 
with four carbon atoms and an impurity site occupied by an oxygen atom 
(furan) or selenium atom (selenophene), and in the case of pyrrole, a 
nitrogen-hydrogen complex.  The effect of the impurity atom in these systems 
is to interrupt the periodicity associated with a ring of carbon atoms by 
replacing the ring with a linear array. In pyrene, the linear region is 
associated with the two interior carbon atoms.  The linear behavior 
of the DPI in pyrrole and similar molecules is characteristic of a coherent 
process where the two photoelectrons are emitted simultaneously with equal 
kinetic energies and oppositely directed momenta \cite{Hube14,Jank14,HubeXX}.  
According to our interpretation, a similar process takes place in pyrene.

\begin{figure}[htb]
\includegraphics[width=8.8cm,clip]{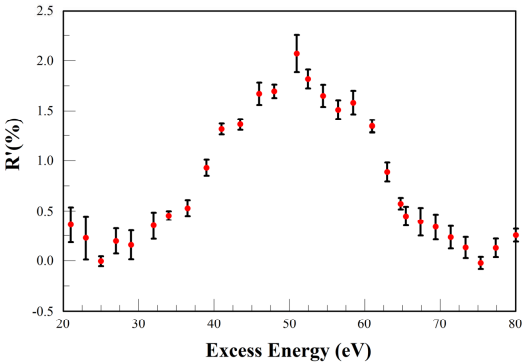}
\caption{
The relative perimeter pairing yield $R'$ in pyrene vs excess energy. The 
data are obtained by subtracting the linear part of the ratio curve $R$ shown 
in Fig. \ref{one} extrapolated down to zero.}
\label{two}
\end{figure}

The mid-range region in pyrene reflects a continuum resonance that is 
likely to be related to the pairing of two mobile electrons in a 
one-dimensional periodic potential \cite{Maha06}. In our analysis, the $\pi$ 
electrons involved in the resonance are associated with the 14 carbon 
atoms on the perimeter of the molecule. In Fig.\ \ref{one}, there is a 
region where the resonance and the linear DPI-mechanism overlap.  
Assuming the contributions from the two processes superpose incoherently, 
we can extrapolate the linear curve to zero, thus obtaining an onset energy of 
61 eV.  Subtracting the linear response from the DPI data we obtain a plot 
of the perimeter resonance structure with an onset at 30 eV, a peak at 
50 eV, and the upper cut-off at 75 eV (Fig.\ \ref{two}).

\section{Coronene}
\begin{figure}[htb]
\includegraphics[width=8.8cm,clip]{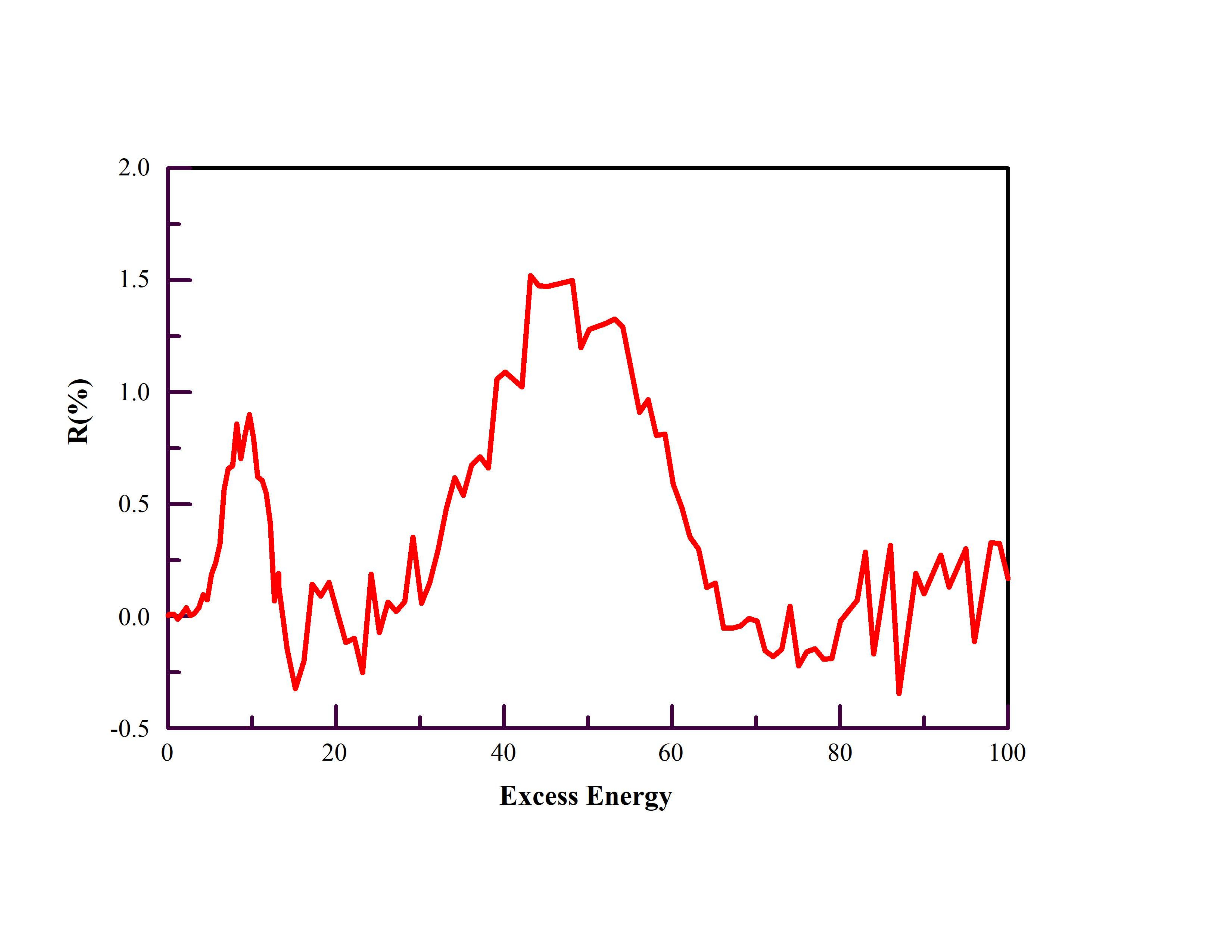}
\caption{
Relative DPI yield $R$ as defined in Eq. (\ref{ratio}) in coronene vs 
excess energy. 
Note the peak at 10 eV and the pairing resonance at higher energy.
}
\label{three}
\end{figure}

It is useful to compare the DPI results for coronene \cite{Hart13} with the 
corresponding pyrene data. Like pyrene, coronene has a 10-eV peak. which 
we attribute to an excitation involving the six carbon atoms in the 
interior.  Unlike pyrene, there is no linear behavior at high energies.  
Instead, coronene has a resonance region, 30 to 70 eV, that
qualitatively resembles what is found for the perimeter resonance in pyrene
(cf.\ Fig.\ \ref{three}). 

\section{Discussion}
\begin{figure}[htb]
\includegraphics[width=8.8cm,clip]{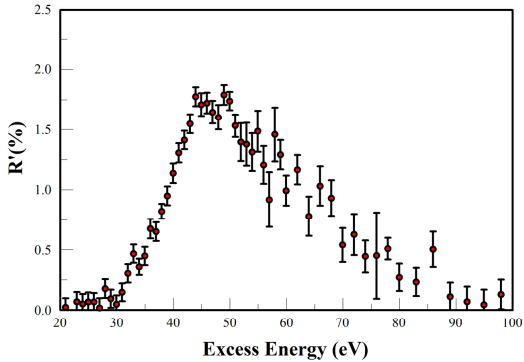}
\caption{
Relative DPI yield $R$ of anthracene \cite{Wehl12} as defined in 
Eq.\ (\ref{ratio}) after subtraction of the linear increase of the ratio 
by a straight line extrapolated down to the abcissa.  
}
\label{five}
\end{figure}

In order to test our model of independent subsystems, we turn now to 
anthracene (C$_{14}$H$_{10}$), which has the same number of perimeter atoms 
as pyrene but no interior atoms. In Fig.\ \ref{five} we display the
reduced resonance peak in anthracene that is obtained after subtracting
the knock-out as well as the linear contribution to the DPI ratio.
This resonance resembles the benzene resonance extending from 30 eV to 
approximately 90 eV above threshold with a peak at ca. 45 eV. It also
resembles the mid-range resonance in pyrene (cf.\ Fig.\ \ref{two}) and
is similar to the mid-range resonance in coronene (cf.\ Fig.\ \ref{three}).
Taken together, these results support the hypothesis of independent DPI
contributions from interior and perimeter carbon atoms.

Finally, we comment on treating the perimeter carbon atoms and the interior 
carbon atoms as independent {\it subsystems}.  The justification for this 
approximation is associated with H\"uckel's Rule, which predicts a high 
stability of an aromatic hydrocarbons for 
\begin{equation}
N = 2 + 4n, n = 0, 1, 2, ...
\end{equation}
where $N$ is the number of $\pi$ electrons (i.e.\ carbon atoms) in the 
planar ring \cite{Huck31}. Although neither pyrene nor coronene satisfy the 
rule when $N$ is the total number of carbon atoms, the rule is satisfied for 
the perimeter subsystem and the interior subsystem in both molecules.  The 
increased stability associated with H\"uckels Rule suggests the subsystems 
are likely to act quasi-independently in DPI processes.

\section{Summary}
We have analyzed the DPI yield in the aromatic hydrocarbons pyrene 
and coronene. In both molecules, there are perimeter and interior carbon 
atoms. The results of our analysis show that the dominant features in the 
yield are independent contributions from the perimeter and interior 
subsystems. It is proposed that the quasi-independence of the two 
subsystems is due to the interior and perimeter subsystems independently 
obeying H\"uckel's Rule. 

\section*{ACKNOWLEDGMENTS}
The authors thank Dr. Narayan Appathurai for critical reading of the 
manuscript and helpful comments,
 

\end{document}